\documentclass[showpacs,preprintnumbers,pre]{revtex4}
\usepackage{amsmath,amssymb,graphicx,subfigure,color}
\newcommand{\average}[1]{\left\langle #1 \right\rangle}
\newcommand{\tauscale}{\tau_{\text{s}}}

\begin{document}
\title{Algebraic Correlation Function and Anomalous Diffusion in the HMF model}
\author{Yoshiyuki Y. Yamaguchi$^{1}$\thanks{E-mail: yyama@amp.i.kyoto-u.ac.jp},
Freddy Bouchet$^{2}$\thanks{E-mail: Freddy.Bouchet@inln.cnrs.fr},
Thierry Dauxois$^3$\thanks{E-mail: Thierry.Dauxois@ens-lyon.fr}}
\affiliation{
1. Department of Applied Mathematics and Physics, Graduate School of Informatics, Kyoto University, 606-8501, Kyoto, Japan.\\
2. Institut Non Lin\'eaire de Nice,  UMR-CNRS 6618, 1361 route des Lucioles 06560 Valbonne, France.\\
3. Laboratoire de Physique, ENS Lyon, CNRS, 46 All\'{e}e d'Italie,
69364 Lyon cedex 07, France}

\date{\today}

\begin{abstract}
In the quasi-stationary states of the Hamiltonian Mean-Field model,
we numerically compute correlation functions of momenta and
diffusion of angles with homogeneous initial conditions. This is an
example, in a N-body Hamiltonian system, of anomalous transport
properties characterized by non exponential relaxations and
long-range temporal correlations. Kinetic theory predicts a striking
transition between weak anomalous diffusion and strong anomalous
diffusion. The numerical results are in excellent agreement with the
quantitative predictions of the anomalous transport exponents.
Noteworthy, also at statistical equilibrium, the system exhibits
long-range temporal correlations: the correlation function is
inversely proportional to time with a logarithmic correction instead
of the usually expected exponential decay, leading to weak anomalous
transport properties.
\end{abstract}
 \maketitle
\section{Introduction}

Recently, a new light was shed on long-range interacting
systems~\cite{Les_Houches}.  The first reason is that a mathematical
characterization~\cite{ellis_turkington} and the study of several
simple models have completely clarified the inequivalence of
ensembles that might exists between the microcanonical and the
canonical ensembles~\cite{barre01,hugo}. The second is the
appearance of a very useful technique, namely the large deviation
theory, to compute the microcanonical number of microstates and thus
the associated microcanonical entropy~\cite{barre-05}.  The third is
a classification of all possible situations of ensemble
inequivalence~\cite{Classification}. The last, but not the least,
reason is the understanding that the broad spectrum of applications
(self-gravitating~\cite{Chavanis_96} and Coulomb systems, vortices
in two-dimensional fluid mechanics, wave-particles interaction,
trapped charged particles, ...)~\cite{Les_Houches} should be
considered simultaneously since significant advances were performed
independently in the different domains. However as usual in Physics,
the study of simple models is of particular interest not only for
pedagogical properties, but also for testing ideas that might be
derived analytically and verified numerically without very expensive
simulations.

We consider here the Hamiltonian Mean Field (HMF) model, which is
considered as the paradigmatic dynamical model for long-range
interacting systems. This model~\cite{zaslavsky_HMF,konishi-kaneko,pichon,antoni-95} consists of $N$ particles moving on the unit
circle, and is described by the Hamiltonian
\begin{equation}
  \label{eq:hamiltonian}
  H = \dfrac{1}{2} \sum_{j=1}^{N} p_{j}^{2}
  + \dfrac{1}{2N} \sum_{j=1}^{N} \sum_{k=1}^{N}
  [ 1-\cos(\theta_{j}-\theta_{k}) ],
\end{equation}
where $\theta_{j}$ is the angle of $j$-th particle and $p_{j}$ its
conjugate momentum. Using a change of the time unit, the prefactor
$1/N$ of the second term is added in order to get an extensive
energy~\cite{barre-05}. Thus, in the limit $N\rightarrow \infty$,
the appropriate mean-field scaling is obtained for the statistical
mechanics. Studies of the HMF model have been recently reinforced by
the discovery of its link with the Colson-Bonifacio model for the
single-pass free electron laser~\cite{barre-05}.

Within this model, a striking disagreement was reported between the
canonical statistical mechanics predictions and time averages of
constant energy molecular dynamics
simulations~\cite{antoni-95,lrt2001}. As the model has only a second
order phase transition~\cite{barre-05} at the critical energy
density $U_{c}=3/4$, the possibility that the origin might lead to
an inequivalence between canonical and microcanonical statistical
mechanics can be excluded~\cite{Classification}. Moreover,
recently, it has been shown unambiguously that the microcanonical
entropy leads to the same predictions than the canonical free
energy~\cite{barre-05}. Very interesting results about the behavior
of such a system in contact with a thermal bath has however been
recently reported~\cite{fulvio1,fulvio2}.

The origin of the apparent disagreement comes from a particularly
slow dynamical evolution of this long-range system. Indeed, in
Hamiltonian systems with long-range interactions, systems are
sometimes trapped in quasi-stationary states (QSS) before going to
equilibrium. Examples of such QSS were found in a 1-dimensional
self-gravitating system~\cite{tsuchiya-96} and in the HMF
model~\cite{yamaguchi-03}. The trapping time diverges algebraically
in the limit $N\to\infty$ and, hence, time averages disagree with
canonical averages if the computing time is not long enough.

To understand the dynamics during such a long period, QSS were
interpreted as stable stationary states of the Vlasov
equation~\cite{yamaguchi-04,barre-06} that can be derived from the
Hamiltonian dynamics. The Vlasov equation, which governs 1-particle
distribution function is indeed exact~\cite{braunhepp} in the limit
$N\to\infty$, but  only approximate for a finite system: finite size
effects drive indeed the system from the Vlasov stable stationary
state to the Boltzmann equilibrium. Recently, Caglioti and
Rousset~\cite{caglioti-04} proved for a wide class of potentials
which includes the HMF case, that $N$ particles starting close to a
Vlasov stable stationary state remain close to it during a time
scale proportional at least to $N^{1/8}$. The result is consistent
with numerical results which state that the lifetime of QSS scales
like $N^{1.7}$~\cite{yamaguchi-03}.

Using a kinetic approach which goes beyond the above Vlasov
interpretation, the correlation function of momenta was recently
derived~\cite{Bouchet-Dauxoisa,Bouchet-Dauxois} with the following
assumptions: (i) a finite but large enough number of particles, (ii)
a homogeneous distribution of angles, and (iii) a system in
a (quasi-)stationary state. As shown in
Refs.~\cite{yamaguchi-04,barre-06}, the latter condition amounts to
consider initial distributions of momenta $f_{0}(p)$ satisfying the
inequality
\begin{equation}
  \label{eq:stability-condition}
   1 + \dfrac{1}{2} \int_{-\infty}^{+\infty}
  \dfrac{f'_{0}(p)}{p} \mbox{d}p > 0.
\end{equation}
This condition has been derived for the linear~\cite{Inagaki} and
formal~\cite{yamaguchi-04}  stability of the distribution $f_0$ (see
also Ref.~\cite{chavanisvattevillebouchet} for another derivation).
This condition defines a critical energy $U_{c}^{\ast}$ which is, in
general, different from the critical energy $U_{c}=3/4$ where the
second order phase transition is located. However, as expected, both
values coincide for a gaussian distribution $f_{0}(p)$. Above theory
is expected to be valid in the time interval $1\ll \tau\ll N$, where
$\tau=t/N$ is the appropriate rescaled time.

Among the main predictions resumed in Table~\ref{tab:Cp}, one might
emphasize that distributions $f_0(p)$ with algebraic tails were
proved to have a correlation function of momenta $C_{p}(\tau)$ with
an algebraic decay in the long-time regime. Striking algebraic large
time behaviors for momentum autocorrelations had  been first
numerically observed in Refs.~\cite{lrt2001,Pluchino}. On the
contrary, distributions with stretched exponential tails correspond
to correlation functions inversely proportional to time with a
logarithmic correction. It is also important to stress that gaussian
distributions, which corresponds to $\delta=2$ in the stretched
exponential case, leads to a long-time correlation of $\ln\tau/\tau$
instead of the usual exponential decay in the stable, supercritical
energy regime $U>U_{c}^{\ast}=U_{c}$, although the initial
distribution is at equilibrium. The origin of this unusual long-time
momentum correlations does not depend on the center part of the
momentum distributions $f_{0}(p)$ but on its tails. One might
understand physically this behavior, by the fact that particles
located in these tails move almost freely, and hence yield long-time
correlations.

\begin{table}[htbp]
  \centering
  \begin{tabular}{|l|c|c|c|}
    \hline
    &&&\\ Tails & $f_{0}(p)$ & $C_{p}(\tau)$
    & $\sigma_{\theta}^{2}(\tau)$\\
    &&&\\
    \hline
    &&&\\Power-law &
    $|p|^{-\nu}$ & $\tau^{-\alpha}$ & $\tau^{2-\alpha}$ \\
    &&&\\
    \hline
    &&&\\
    Stretched exponential &
    $\quad\exp(-\beta|p|^{\delta})$\quad & \quad$\dfrac{(\ln\tau)^{2/\delta}}{\tau}$\quad
    & \quad$\tau (\ln\tau)^{2/\delta+1}$\quad\\
    &&&\\\hline
  \end{tabular}
  \caption{Asymptotic forms of initial distributions $f_{0}(p)$,
    and theoretical predictions
    of correlation functions $C_{p}(\tau)$
    and the diffusion $\sigma_{\theta}^{2}(\tau)$ in the long-time regime.
    Asymptotic forms of the distribution
    and the predictions are assumed and predicted
    in the limits  $|p|\to\infty$ and $\tau\to\infty$ respectively,
    where $\tau=t/N$ is a rescaled time.
    The exponent $\alpha$ is given as
    $\alpha=(\nu-3)/(\nu+2)$. See Ref.~\cite{Bouchet-Dauxois} for details.}
  \label{tab:Cp}
\end{table}

In these (quasi-)stationary states, the theoretical law for the
diffusion of angles $\sigma_{\theta}^{2}(\tau)$ has been also
derived. The predictions~\cite{Freddy_PRE,Bouchet-Dauxois} for the
diffusion properties are listed in Table~\ref{tab:Cp}. They clarify
the highly debated disagreement between different numerical
simulations reporting either anomalous~\cite{latora-99} or
normal~\cite{yamaguchi-03} diffusion, in particular by delineating
the time regime for which such anomalous behavior should occur. We
briefly recall that when the moment of order $n$ of the distribution
scales like $\tau^{n/2}$ at large time, such a transport is called
{\em normal}. However, {\em anomalous}
transport~\cite{bouchaud,majda,Castiglione}, where moments do not
scale as in the diffusive case, were reported in some stochastic
models, in continuous time random walks (Levy walks), and for
systems with a lack of stationarity of the corresponding stochastic
process~\cite{Fermi}. When the distribution $f_0(p)$ is changed
within the HMF model, a transition between weak anomalous diffusion
(normal diffusion with logarithmic corrections) and strong anomalous
diffusion is thus predicted.  From the physical point of view, as
particles with large momentum $p$ fly very fast in comparison to the
typical time scales of the fluctuations of the potential, they are
subjected to a very weak diffusion and thus maintain their large
momentum during a very long time. A thick distribution of waiting
time with a large momentum explains the anomalous diffusion. From a
mathematical point of view, these behaviors are linked to the non
exponential relaxation of the Fokker-Planck equation describing the
diffusion of momenta, leading to long-range temporal
correlations~\cite{Bouchet-Dauxois}.  This mechanism is new in the
context of kinetic theory. However, similar Fokker-Planck equations,
with a rapidly vanishing diffusion coefficients obtained by other
physical mechanisms, have been studied in several
frameworks~\cite{farago,micciche,lutz}.

The first purpose of this article is to numerically check the
theoretical predicted correlation functions for power-tail and
gaussian distributions by using accurate numerical simulations. The
other is to clarify whether diffusion is normal or anomalous, which
depends on the choice of the initial distribution $f_0(p)$.

The article is organized as follows. Some useful quantities are
introduced in Section~\ref{sec:quantities}. In
Sections~\ref{sec:power-tail} and \ref{sec:gaussian}, we
respectively focus on initial distributions with power-law and
gaussian tails. In each section, we first check the stationarity and
the stability following the method developed in
Ref.~\cite{yamaguchi-04} and determine the time region of the QSS.
We also study carefully the correlation function and the diffusion
comparing them with theoretical predictions. Finally,
section~\ref{sec:summary} concludes the discussion.

\section{Quantities of interest and numerical protocol}
\label{sec:quantities}

In order to check the stationarity and the stability of an initial
distribution $f_{0}(p)$, we study the temporal evolutions of several
macrovariables:\begin{itemize}
                 \item the magnetization defined
as the modulus $M$ of the vector ${\bf M}=(M_{x},M_{y})$, where both
components are defined as
 $ M_{x} = \average{\cos\theta}_{N}$ and $ M_{y} = \average{\sin\theta}_{N}
$. The bracket $\average{\cdot}_{N}$ represents the average over all
 particles, for instance $\average{\cos\theta}_{N} =(
\sum_{j=1}^{N} \cos\theta_{j})/N$. Note that the magnetization $M$
is constant if the system is stable stationary.
                 \item the moments of the $1$-body
distribution function $f(\theta,p,t)$. As explained in details in
Ref.~\cite{yamaguchi-04}, the stationarity of the $1$-body
distribution $f(\theta,p,t)$ implies the stationarity of the
individual energy distribution $f_{e}(e,t)$, where
$e=p^{2}/2-M_{x}\cos\theta-M_{y}\sin\theta$. Moreover, the
stationarity of $f_{e}(e,t)$ implies the stationarity of all moments
$\mu_{n}=\average{e^{n}}_{N}$. As the stationarity of the moment is
a necessary condition for stability, vanishing derivatives
$\dot{\mu}_{n}=\mbox{d}\mu_{n}/\mbox{d}t$, for $n=1,2$ and $3$,
would suggest that the system is in a (quasi-)stationary state,
while large derivatives clearly indicate a non-stationary state. In
addition, the stability is suggested if a state stays stationary for
a long period.
\end{itemize}

While checking the stationarity and the stability, we identify a
time region where the system is in the QSS, during which we observe
the correlation function of momenta
$C_{p}(\tau)=\average{p(\tau)p(0)}_{N} $ and the diffusion of angles
$
\sigma_{\theta}^{2}(\tau)=\average{[\theta(\tau)-\theta(0)]^{2}}_{N}.
$  The latter quantity  can be rewritten as follows
\begin{equation}
  \dfrac{\sigma_{\theta}^{2}(\tau)}{N^{2}}
  = \int_{0}^{\tau} \mbox{d}\tau_{1} \int_{0}^{\tau} \mbox{d}\tau_{2}
  \average{p(\tau_{1})p(\tau_{2})}_{N}
  = 2 \int_{0}^{\tau} \mbox{d}s \int_{0}^{\tau-s} \mbox{d}\tau_{2}
  \average{p(s+\tau_{2})p(\tau_{2})}_{N},
\end{equation}
where the factor $1/N^{2}$ comes from the time rescaling $\tau=t/N$,
while the new variable $s=\tau_{1}-\tau_{2}$ was introduced to take
advantage of the division of the square domain into  two isoscale
triangles corresponding to $s>0$ and $s<0$. In the
(quasi-)stationary states, the integrand
$\average{p(s+\tau_{2})p(\tau_{2})}_{N}$ does not depend on
$\tau_{2}$  (the QSS evolve on a time scale much larger than $N$)
and hence diffusion can be simplified~\cite{yamaguchi-03} by using
the correlation function  as
\begin{equation}
  \label{eq:sigma_Cp}
  \dfrac{\sigma^{2}_{\theta}(\tau)}{N^{2}}
  = 2 \int_{0}^{\tau} (\tau-s) C_{p}(s)~\mbox{d}s.
\end{equation}

We numerically performed the temporal evolution of the canonical
equations of motion by using a $4$-th order symplectic integrator
\cite{yoshida-90,yoshida-93} with a time step $\Delta t=0.2$ and a
total momentum set to zero.
Initial values of angles being
randomly chosen from a homogeneous distribution, the magnetization
$M$ is hence of order $1/\sqrt{N}$. Omitting this vanishing value of
$M$, the energy density $U= K + ({1-M^{2}})/{2}$ where $K$ is the
kinetic energy density can thus be well approximated by the kinetic
energy density $K$ as $U=K+1/2$. To characterize the simulations,
the only remaining point is the initial distribution of momenta: in
the following sections, as anticipated, we will carefully study
distributions with power-law and gaussian tails.

\section{Power-law tails}
\label{sec:power-tail}

\subsection{Initial distribution}
In this section, we consider the initial distribution
\begin{equation}
  f_{0}(p) = \dfrac{A_{\nu}}{1+|p/p_{0}|^{\nu}},
\end{equation}
whose power-law tails are characterized by the exponent~$\nu$. The
unity, added in the denominator to avoid the divergence at the
origin $p=0$, does not affect neither the asymptotic form, nor the
theoretical predictions. The parameter $p_{0}$ is directly
determined by the kinetic energy density as $ p_{0} = \left(
{2K\sin(3\pi/\nu)}/{\sin(\pi/\nu)} \right)^{1/2}$, while the
normalization factor is
\begin{equation}
  A_{\nu} = \dfrac{\nu}{2\pi} \left(
    \dfrac{\sin^{3}(\pi/\nu)}{2K\sin(3\pi/\nu)} \right)^{1/2} .
\end{equation}
From the stability criterion (\ref{eq:stability-condition}), one
gets that this initial state is Vlasov stable when the kinetic
energy density satisfies the condition
\begin{equation}
  K > \dfrac{1}{4} \dfrac{\sin(\pi/\nu)}{\sin(3\pi/\nu)}.
\end{equation}
 One thus gets a dynamical critical energy
$U_{c}^{\ast}=0.75,~0.625$ and $0.60355\dots$ for $\nu=4,~6$ and $8$
respectively. In the rest of this section, we set the exponent $\nu$ to $8$.

\subsection{Stationarity and stability checks}
\label{sec:power-stationary} 
\begin{figure}[htbp]
  \centering
  \subfigure[Unstable ($U=0.6$)]
  {\includegraphics[width=6.5cm]{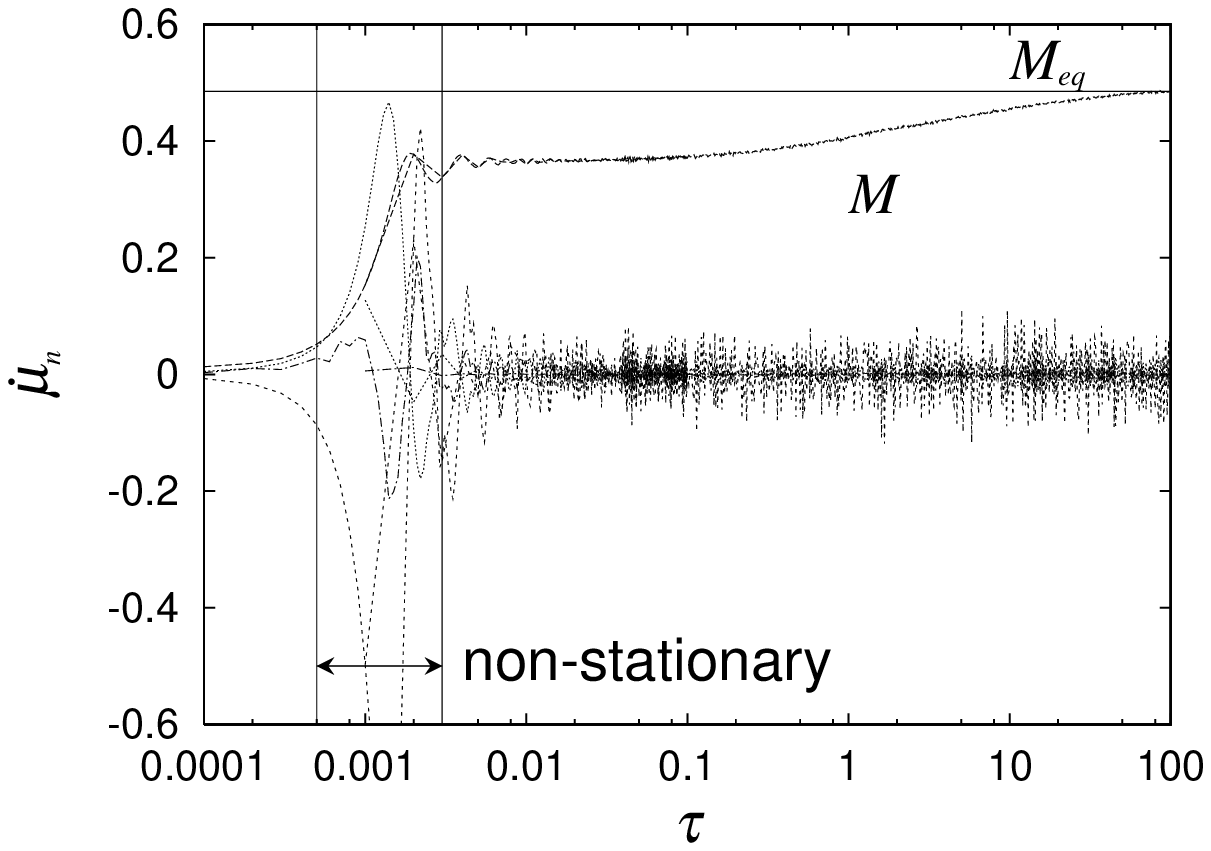}}
  \subfigure[Stable ($U=0.7$)]
  {\includegraphics[width=6.5cm]{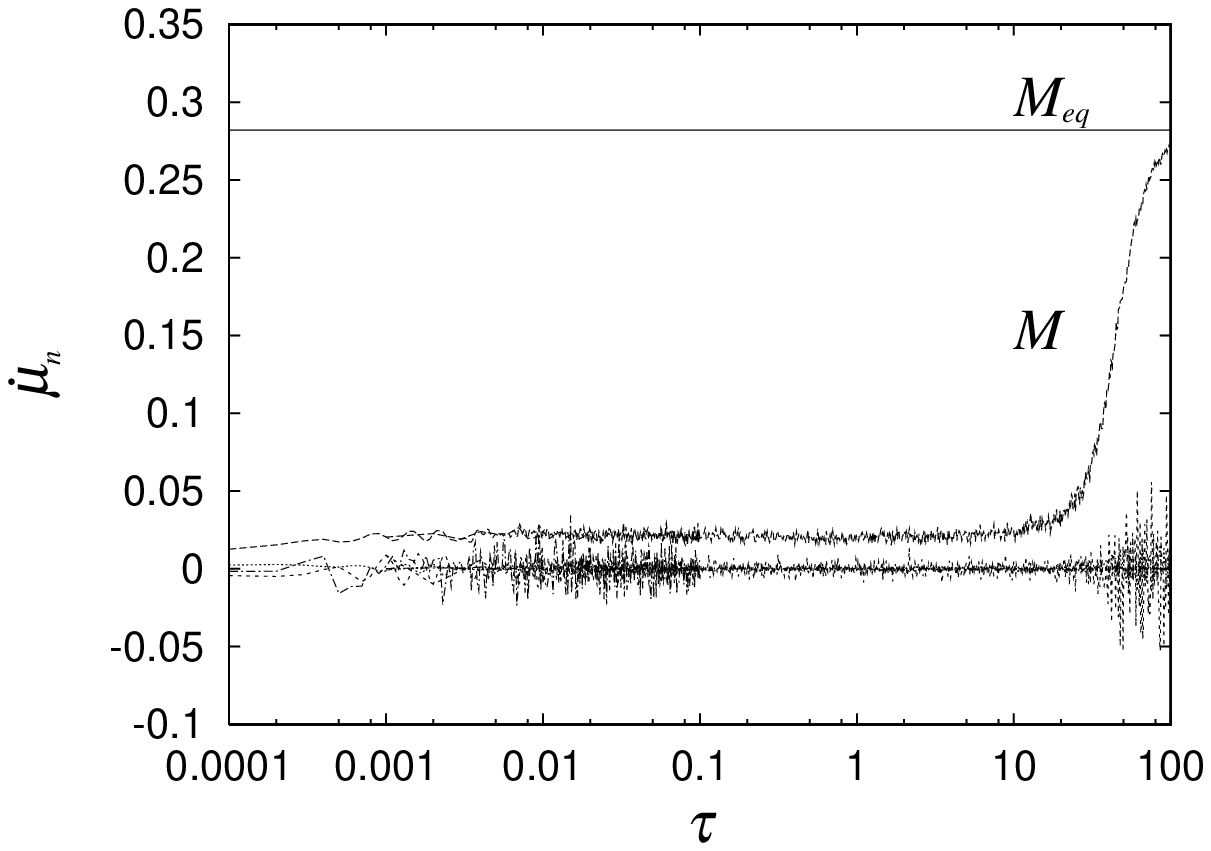}}
  \caption{Stationarity check for initial distributions with
  power-law tails. Note the logarithmic scale for the rescaled time $\tau=t/N$.
    Panel (a) presents the unstable case $U=0.6$ while panel (b)
    the stable one $U=0.7$.
    The three curves  $\dot{\mu}_{n}~(n=1,2,3)$ are reported
    in both panels. Their vertical scales are multiplied
    by $100$ for graphical purposes.
    Curves and horizontal lines indicated by symbols $M$ and $M_{eq}$
    represent respectively temporal evolutions of the magnetization
    and its equilibrium value. All numerical curves are obtained
    by averaging $20$ different numerical simulations for $N=10^{4}$.}
  \label{fig:stationarity-power}
\end{figure}

Let us numerically check the
stationarity and the stability of these states; in particular, it
will clarify the time region of existence of the QSS.
Figure~\ref{fig:stationarity-power} presents the temporal evolution
of $M$ and $\dot{\mu}_{n}~(n=1,2,3)$ for the unstable
($U=0.6<U_{c}^{\ast}$) and stable ($U=0.7>U_{c}^{\ast}$) cases. In
both cases, the magnetization $M$ eventually goes toward the
equilibrium value $M_{eq}$, indicated by horizontal lines. The three
quantities $\dot{\mu}_{n}$ have vanishing small fluctuations around
zero except during the time interval $0.0005<\tau<0.003$ for the
unstable case. In the unstable case, the system is first in an
unstable stationary state ($\tau<0.0005$), before becoming
non-stationary ($0.0005<\tau<0.003$) and finally reaches stable
stationary states ($\tau>0.003$). On the other hand, in the stable
case, the stable stationarity holds throughout the computed time.

In the stable case, the magnetization $M$ stays around zero before
taking off  around $\tau=20$ to reach the equilibrium value
$M_{eq}$. The fluctuation level of $\dot{\mu}_{n}$ increases around
the take-off time $\tau=20$, but the increase does not imply any
non-stationarity of the system, since the fluctuation level is $10$
times smaller than the corresponding one in the non-stationary time
region of the unstable case. The nonzero magnetization might be at
the origin of the larger fluctuations than in the zero magnetization
cases, since the former has a phase and an individual energy $e$
which depends not only on the modulus $M$ but also on the phase.

\subsection{Check of the theoretical prediction}
\label{sec:power-theory}

\begin{figure}[htbp]
  \centering
  \subfigure[Magnetization]
  {\includegraphics[width=6.5cm]{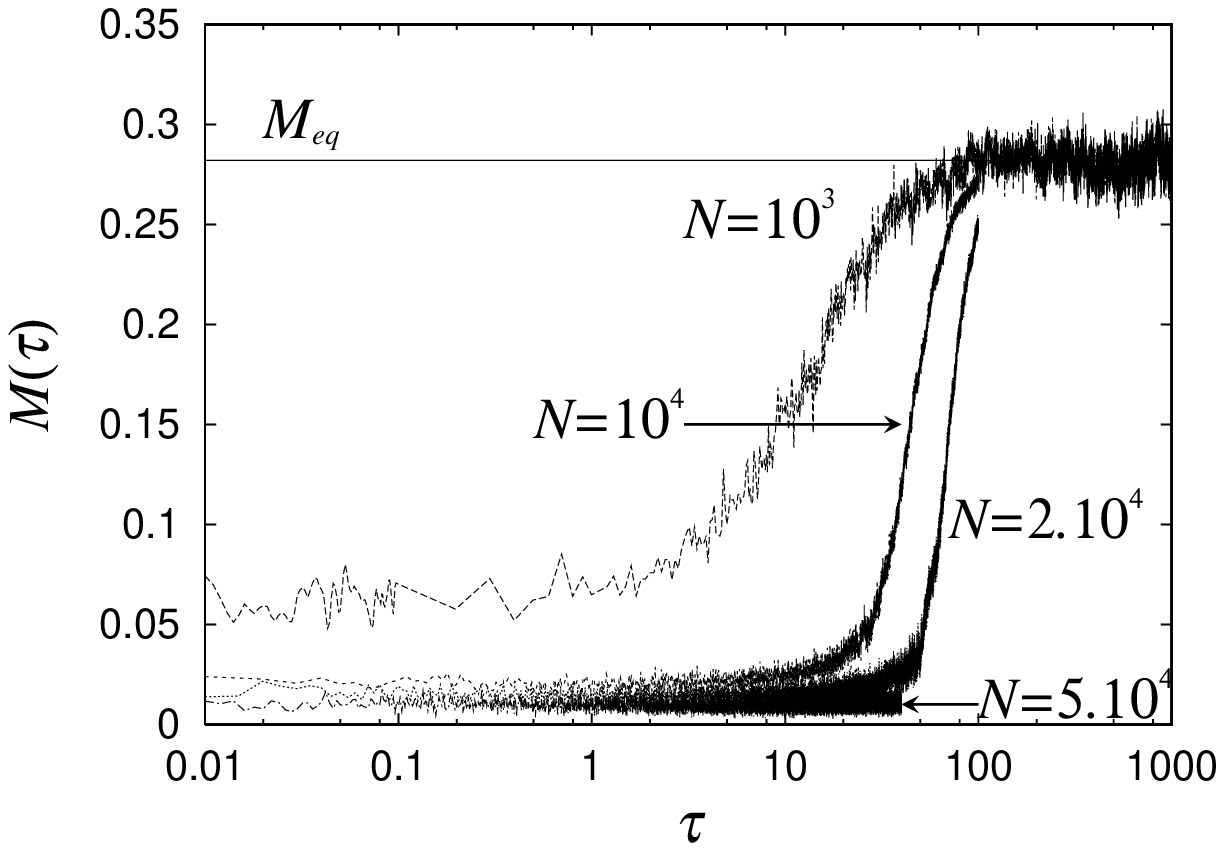}}
  \subfigure[Correlation]
  {\includegraphics[width=6.5cm]{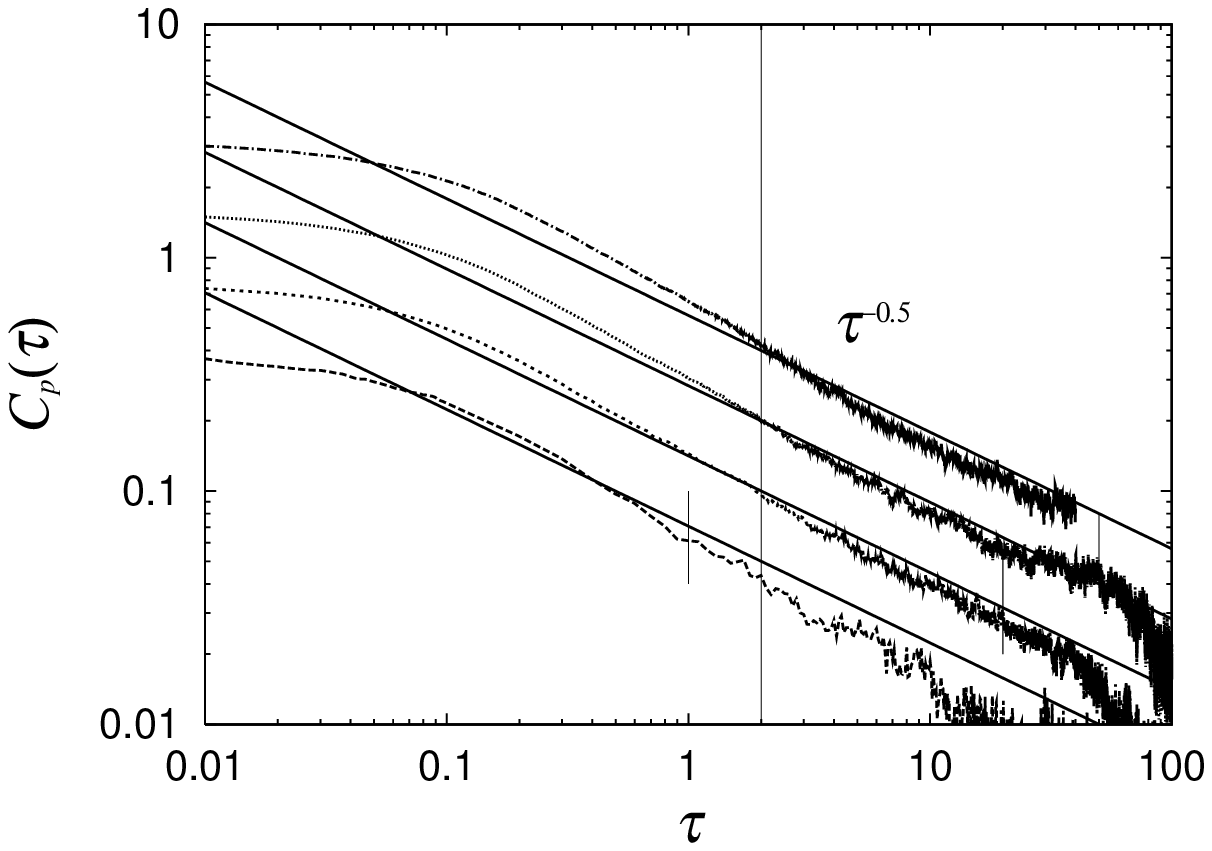}}
  \subfigure[Diffusion]
  {\includegraphics[width=6.5cm]{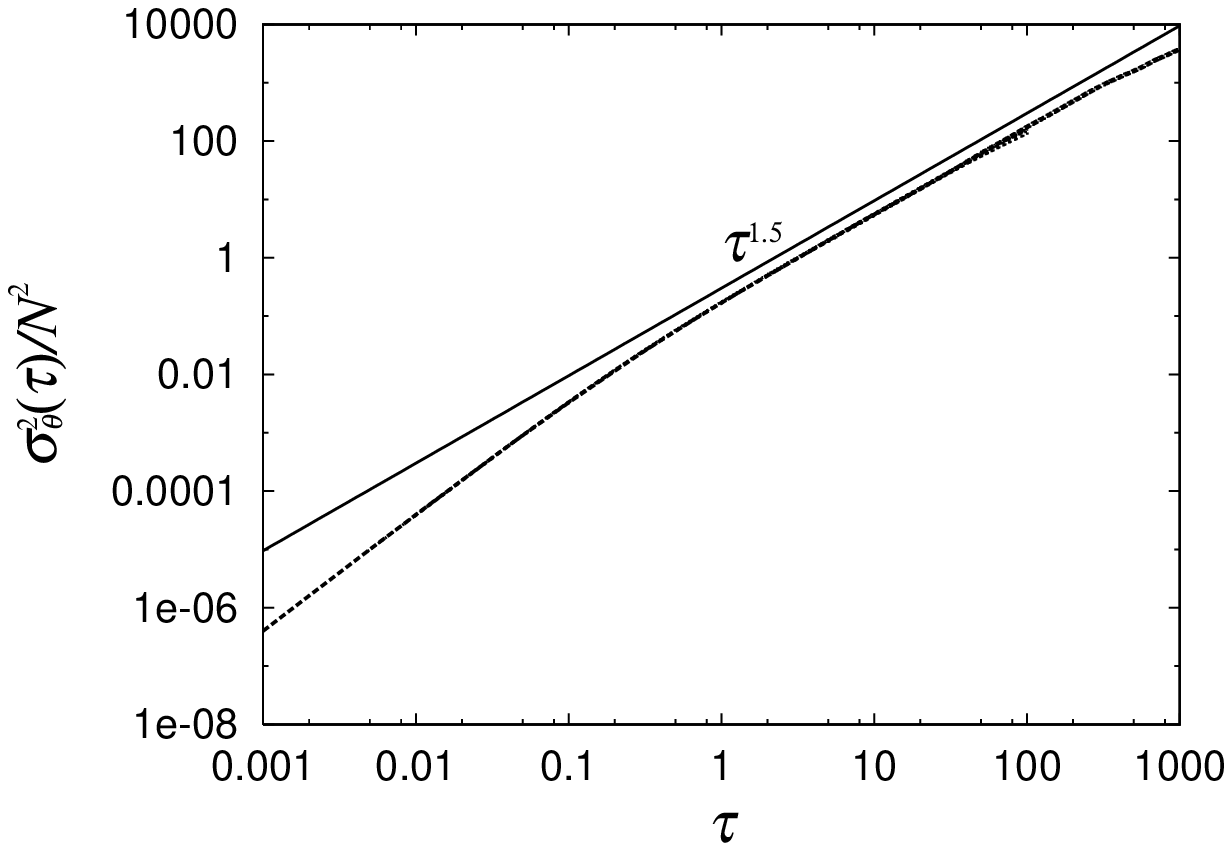}}
  \subfigure[Exponent]
  {\includegraphics[width=6.5cm]{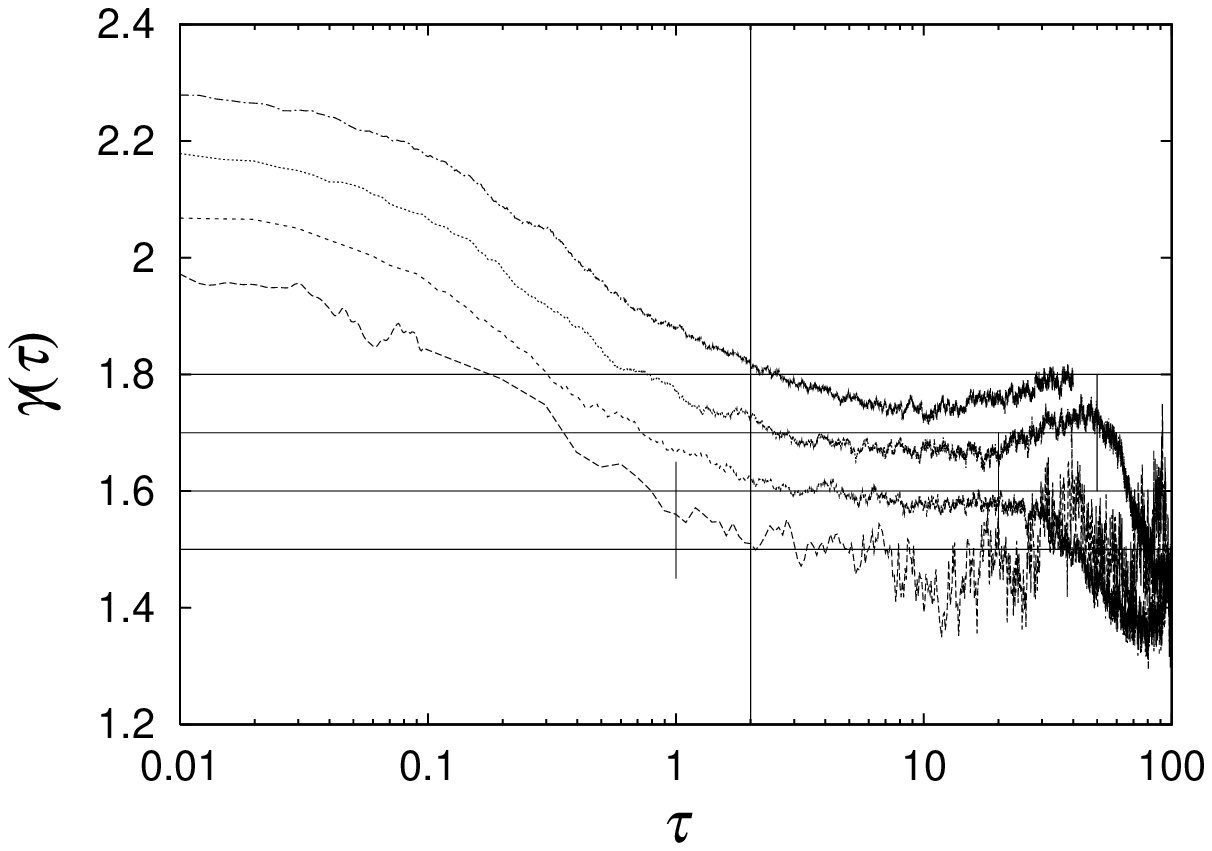}}
  \caption{Check of the theoretical prediction for stable initial
    distributions with power-law tails, in the case $U=0.7$.
    Points are numerically obtained by averaging $20,20,10$ and $5$
    realizations for $N=10^{3},10^{4},2.10^{4}$ and $5.10^{4}$ respectively.
    Panel (a) shows temporal evolution of magnetization
    in the scale time $\tau=t/N$.
    The take-off times of $M$ are estimated as $\tau_{2}=1,20$ and $50$
    for $N=10^{3},10^{4}$ and $2.10^{4}$ respectively,
    and $\tau_{2}$ are marked in panel (b) and (d).
    No take-off for $N=5.10^{4}$ is observed in this computing time.
    The horizontal line represents the equilibrium value of $M$.
    In panel (b), four curves represent the correlation functions of momenta,
    while the straight lines with the slope $-1/2$
    represent the theoretical prediction.
    The curves and the lines are multiplied from the original
    vertical values by $2,4$ and $8$ for $N=10^{4},2.10^{4}$
    and $5.10^{4}$ for graphical purposes.
    The vertical line indicates $\tau_{1}=2$
      from which the valid time region of the theory starts.
    Similarly, panel (c) presents the diffusion of angles,
    while the straight line with the slope $3/2$
    is theoretically predicted.
    The four curves for the four different values of $N$
      are reported and almost collapse.
    Finally, panel (d) shows the temporal evolution of the
    instantaneous exponent $\gamma$, defined in Eq.~(\ref{instantexponent}),
    and $\gamma$ stays around the theoretically predicted value $3/2$
    in the time region $\tau_{1}<\tau<\tau_{2}$. 
    The values $0.1,0.2$ and $0.3$ are added
    in vertical values for $N=10^{4},2.10^{4}$ and $5.10^{4}$
    respectively for graphical purposes.}
  \label{fig:Cp-power}
\end{figure}

In the stable case ($U=0.7$), we perform numerical
computations for $N=10^{3},10^{4},2.10^{4}$ and $5.10^{4}$, and
average over 20, 20, 10 and $5$ sample orbits respectively. Temporal
evolutions of magnetization $M$ are shown in
Fig.~\ref{fig:Cp-power}(a), and $M$ takes off toward the equilibrium
value $M_{eq}$ around $\tau_{2}=1,20$ and $50$ for $N=10^{3},10^{4}$
and $2.10^{4}$ respectively. The take-off time defines the end of
applicable time region of the theory since the homogeneous
assumption (ii) breaks. Note that no take-off time appears in the
case $N=5.10^{4}$, within the computed time interval.

The theory predicts (see Table~\ref{tab:Cp} for $\nu=8$) that the
correlation function decays algebraically with the exponent $-1/2$,
i.e. $C_{p}(\tau)\sim\tau^{-1/2}$, up to the take-off time
$\tau_{2}$. According to Fig.~\ref{fig:Cp-power}(b), the theoretical
prediction agrees well with numerical computations in the
intermediate time region $\tau_{1}<\tau<\tau_{2}$, where
$\tau_{1}=2$ for any value $N$. This is expected since, on the one
hand, the short-time region $\tau<\tau_{1}$ is out of the time
domain of application since the theory gives asymptotic estimates.
The time $\tau_{1}$ is marked as a long vertical line in
Fig.~\ref{fig:Cp-power}(b) to clearly indicate the start of the
applicable time domain. Although the quantity $\tauscale\simeq
0.005$ is not derived theoretically, the straight lines with the
slope $-1/2$ in Fig.~\ref{fig:Cp-power}(b), representing $( \tau /
\tauscale)^{-1/2}$, emphasizes the agreement of the predicted
exponent.

Introducing the expression of  the correlation function in relation
(\ref{eq:sigma_Cp}) leads to the law
$\sigma_{\theta}^{2}(\tau)\sim\tau^{3/2}$:
Figure~\ref{fig:Cp-power}(c), in which the four
curves for the four different values of $N$ almost collapse,
attests also the validity of this prediction in the intermediate
time region $\tau_{1}<\tau<\tau_{2}$. It is possible to confirm more
precisely that the diffusion exponent is $3/2$ by
introducing the instantaneous exponent $\gamma$~\cite{moyano-06}
defined as
\begin{equation}
  \gamma
  = \dfrac{\mbox{d}\ln\sigma_{\theta}^{2}(\tau)}{\mbox{d}\ln\tau}
  = \dfrac{1}{\sigma_{\theta}^{2}(\tau)}
  \dfrac{\mbox{d}\sigma_{\theta}^{2}(\tau)}{\mbox{d}\ln\tau}
  .\label{instantexponent}
\end{equation}
The instantaneous exponent $\gamma$, shown in
Fig.~\ref{fig:Cp-power}(d), goes down and once
crosses $3/2$. However, $\gamma$ comes back and stays around $3/2$
in the time interval $\tau_{1}<\tau<\tau_{2}$. Above result
confirms therefore unambiguously that the diffusion is anomalous,
namely superdiffusive, in the intermediate QSS time interval as
predicted by the theory~\cite{Bouchet-Dauxois}.

The temporal evolution of $\gamma$ was also recently
discussed by Antoniazzi et al.~\cite{antoniazzi-06}, and was shown
to monotonically decrease toward 1. The difference has two different
origins: First, Antoniazzi et al considered non-homogeneous initial
distribution of angles, which are out of the applicable range of the
theory tested here. Second they considered a waterbag initial
distribution of momenta, which does not have tails initially,
although tails develop of course as soon as the time is slightly
positive. As the theory states that the asymptotic law for diffusion
is determined by the tails of the initial distribution of momenta,
there is no contradiction that the temporal evolution of $\gamma$ is
different. A similar remark applied with the out-of-equilibrium
initial distribution discussed  by Moyano and
Anteneodo~\cite{moyano-06}.

For the power-tail initial distributions, the theoretical
predictions are essentially good, but not exact. We first note that
the increase of $N$ does not affect neither the correlation
function, nor the diffusion, at least for $10^{4}\leq N\leq
5.10^{4}$ (the case $N=10^{3}$ has been excluded since no validity
time region $\tau_{1}<\tau<\tau_{2}$ appears). In the numerical
results, the slope of the diffusion $\gamma$ is not $1.5$ but
belongs to $[1.44, 1.48]$. The relative discrepancy is thus at most
of 4 percents. There are two possibilities to understand this small
discrepancy: (a) the lack of the samples, or (b) the lack of
stationarity which is the assumption (iii) of the theory. We will
discuss on the origin of these discrepancies at the end of the next
section.

\section{Gaussian distribution}
\label{sec:gaussian}

\subsection{Initial distribution}
In this section, we consider the gaussian initial distribution
\begin{equation}
  f_{0}(p) = \dfrac{1}{\sqrt{2\pi T}} e^{-p^{2}/2T},
\end{equation}
where the initial temperature $T$ is determined from the energy
density as $T=2K=2U-1$. The dynamical critical energy of this
gaussian distribution coincides with the critical energy of the
second order phase transition $U_{c}=3/4$. As the distribution of
angles is homogeneous, the system is therefore at equilibrium for
any supercritical energy $U>U_{c}$.

\subsection{Stationarity and stability checks}
\begin{figure}[htbp]
  \centering
  \subfigure[Unstable ($U=0.7$)]{
    \includegraphics[width=6.3cm]{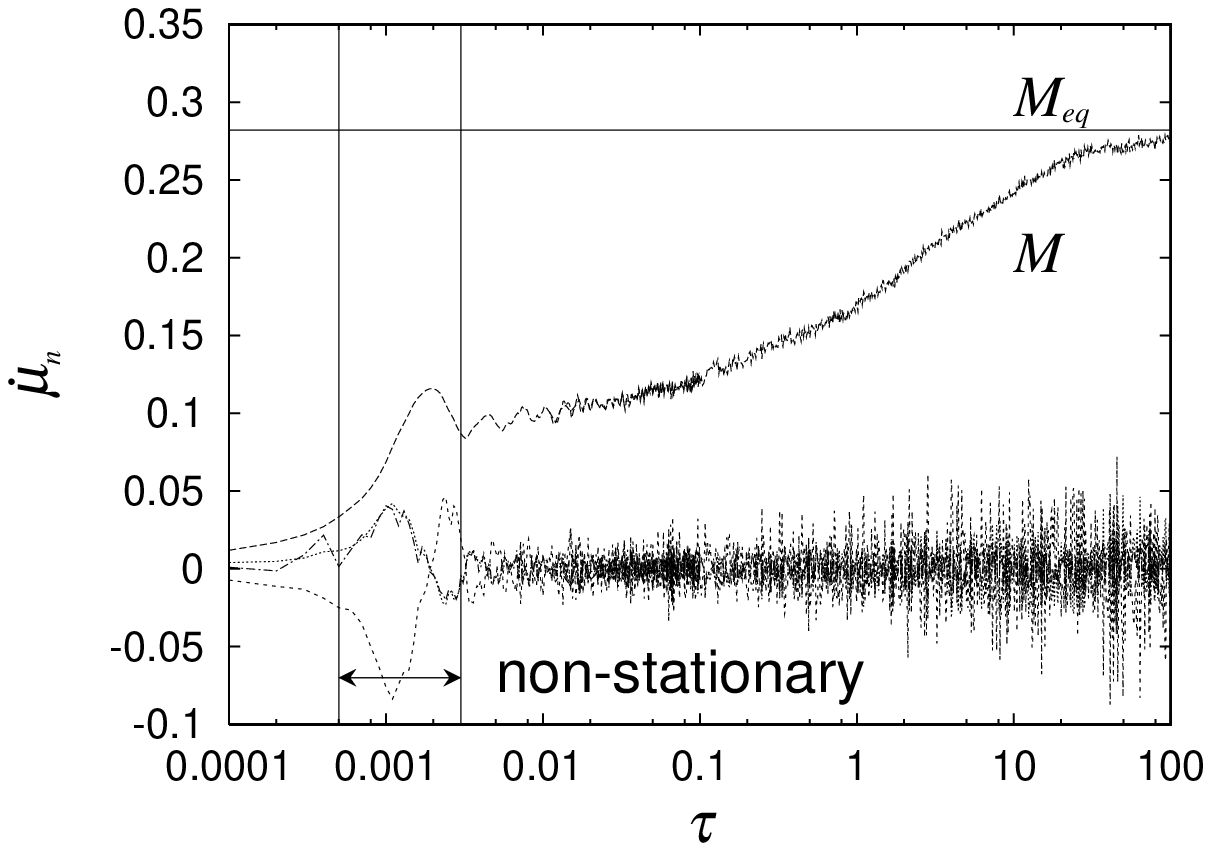}}
  \subfigure[Stable ($U=0.8$)]{
    \includegraphics[width=6.3cm]{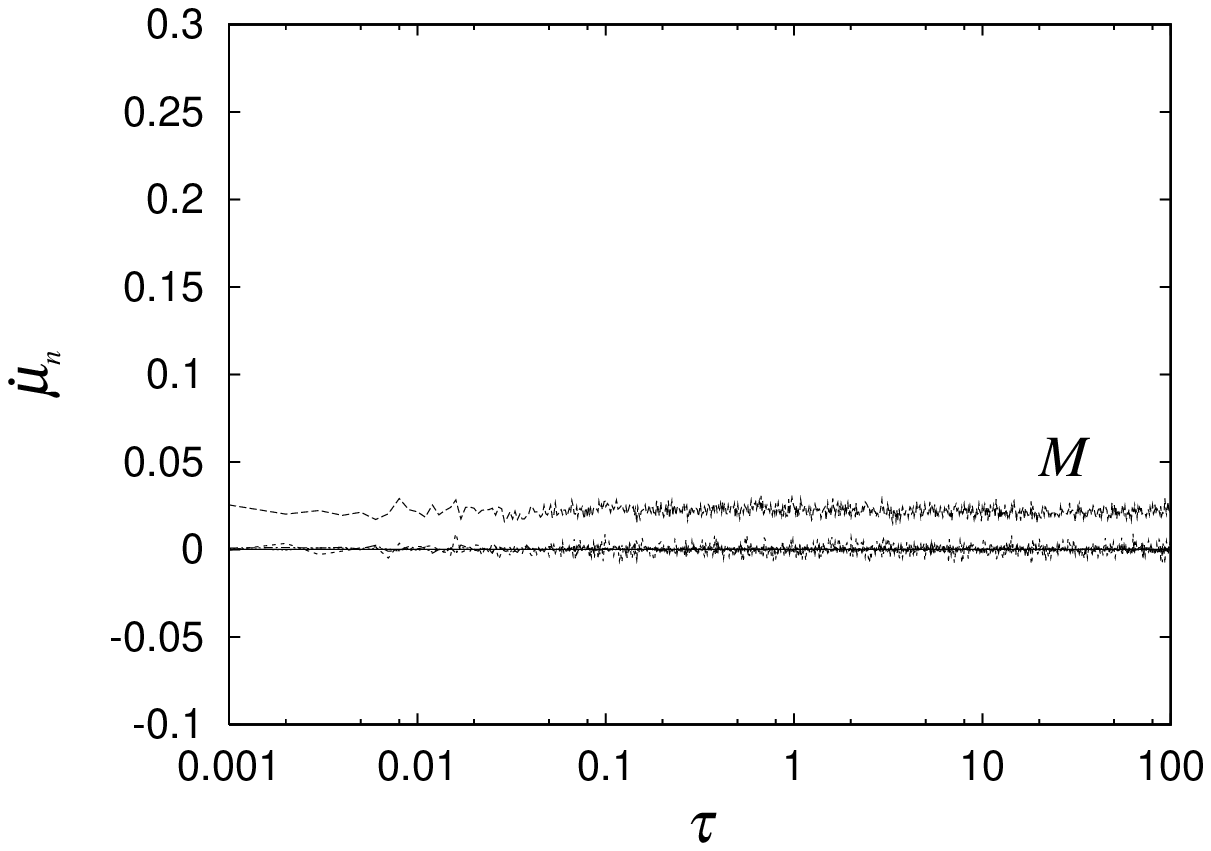}}
  \caption{Stationarity check for gaussian initial distributions
    with $N=10^{4}$.
    Note the logarithmic scale for the rescaled time $\tau=t/N$.
    Panel~(a) presents the unstable case $U=0.7$ while panel (b) the
    stable one $U=0.8$.  The three curves $\dot{\mu}_{n}~(n=1,2,3)$
    are reported in both panels. Their vertical scales are multiplied
    by $100$ for graphical purposes.  Curves indicated by symbol $M$
    represent the temporal evolutions of the magnetization. In panel
    (a), the horizontal line indicated by $M_{eq}$ represents the
    equilibrium value, while in panel (b), the equilibrium value is
    zero.  All numerical curves are obtained by averaging $20$
    different numerical simulations.}
  \label{fig:stationary-gaussian}
\end{figure}

The stationarity and stability are checked as in
Sec.~\ref{sec:power-stationary} by considering temporal evolutions
of magnetization and the derivatives of moments $\mu_n$ shown in
Fig.~\ref{fig:stationary-gaussian}. The scenario of relaxation of
this initial distribution with power-law tails is very similar. In
the unstable case $(U=0.7<U_{c})$, the system reaches stable
stationary states after experiencing unstable stationary and
non-stationary states. In the stable case $(U=0.8>U_{c})$, the state
is stable stationary in the whole time domain since it is initially
at equilibrium.

\subsection{Check of the theoretical prediction}
\begin{figure}[htbp]
  \centering
  \subfigure[Correlation]{
    \includegraphics[width=6.3cm]{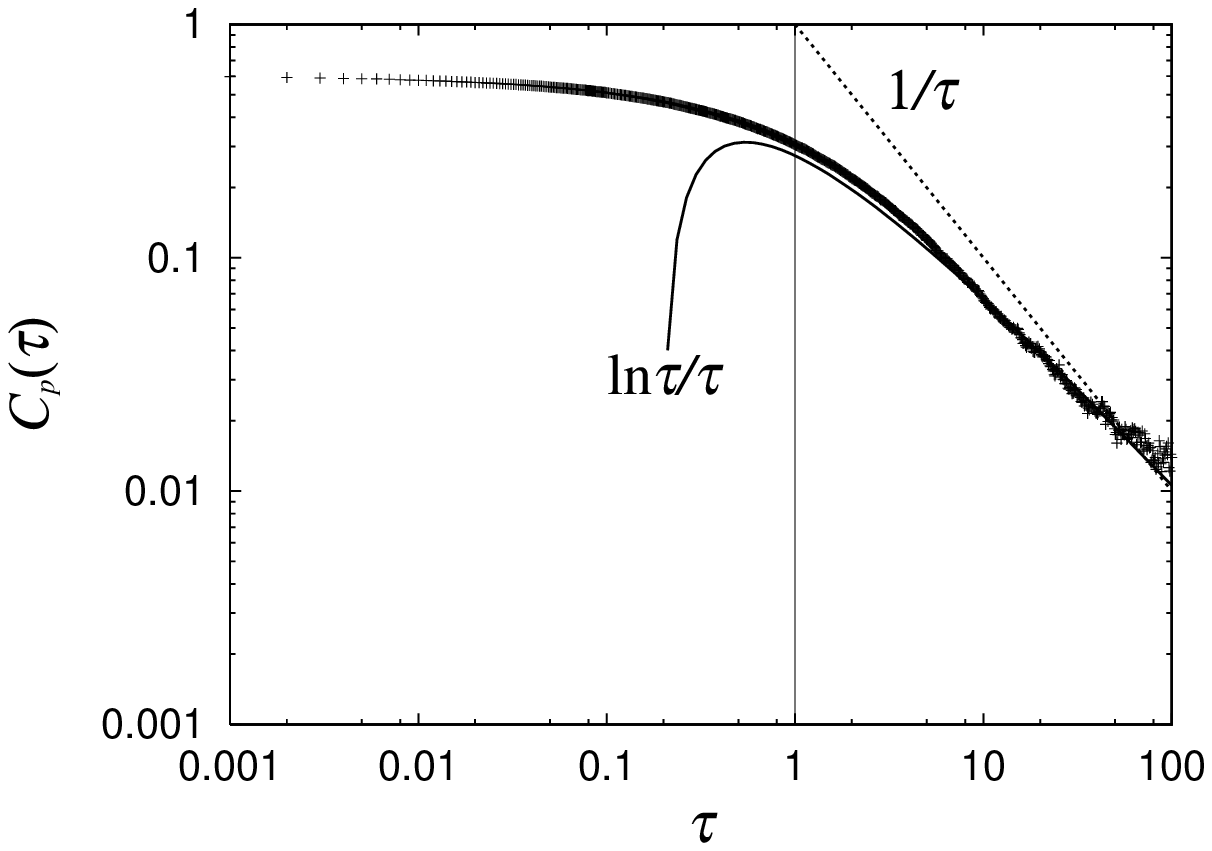}}
  \subfigure[Diffusion]{
    \includegraphics[width=6.3cm]{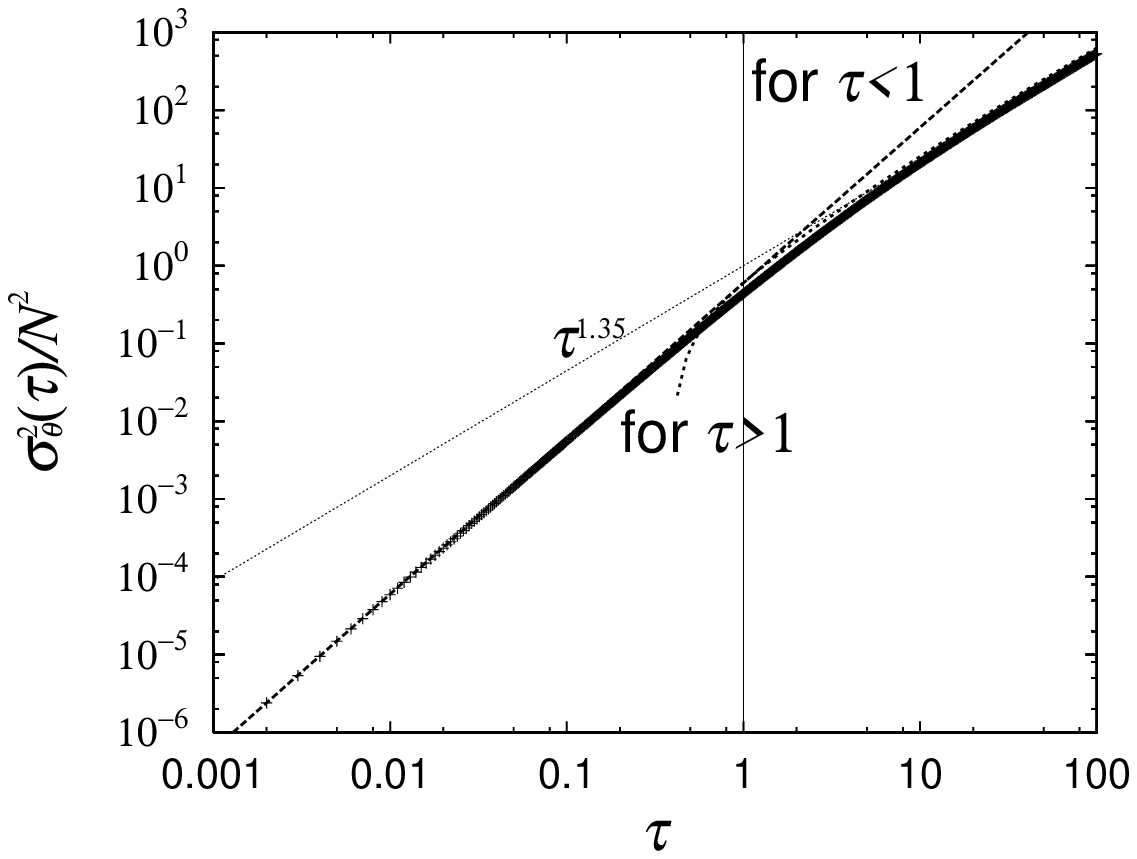}}
  \subfigure[Logarithmic correction]{
    \includegraphics[width=6.3cm]{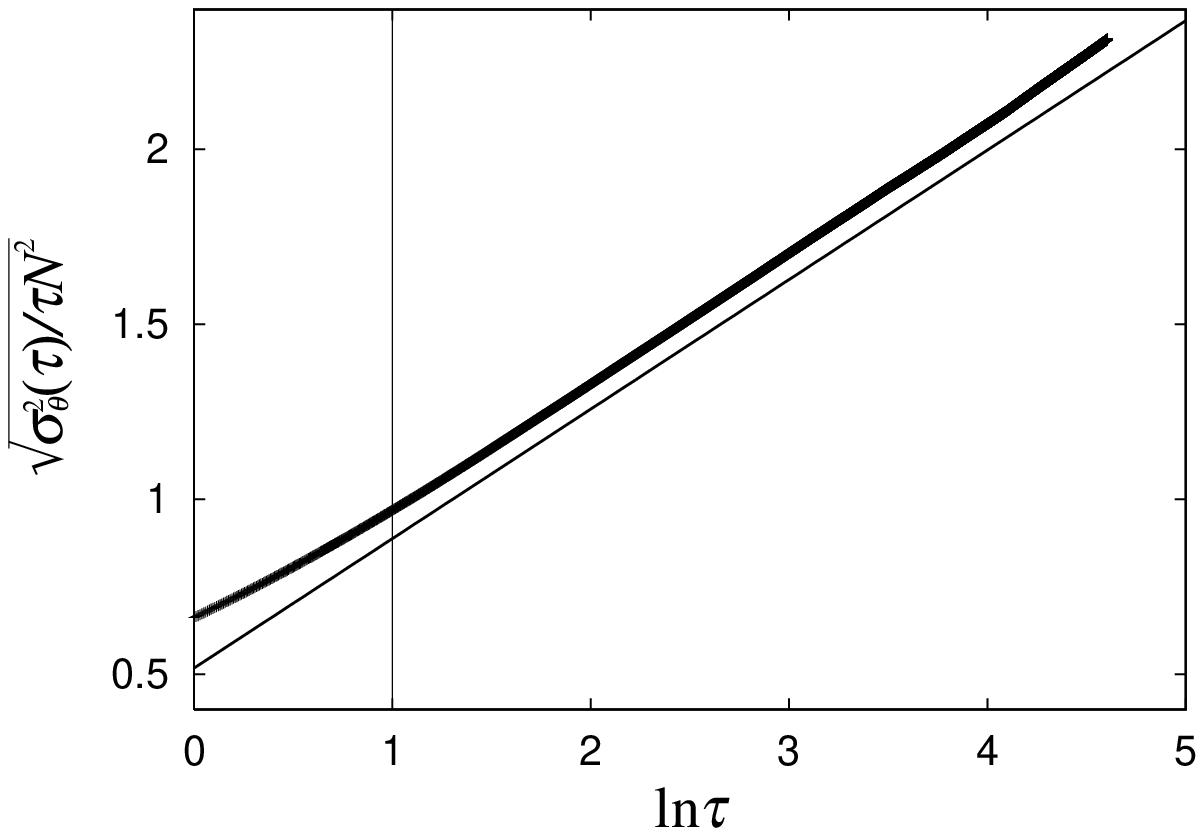}}
  \subfigure[Exponent]
  {\includegraphics[width=6.5cm]{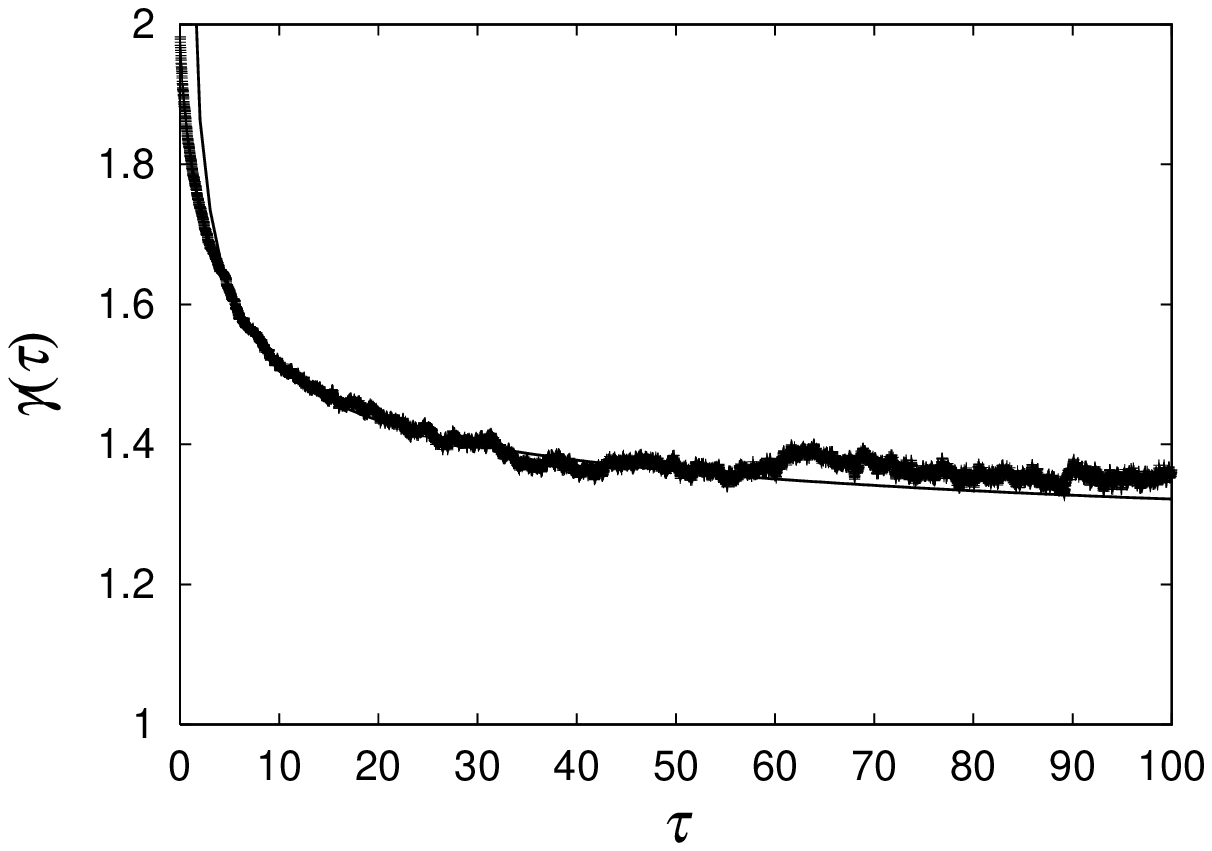}}
  \caption{ Check of the theoretical prediction for gaussian
    stable initial distributions in the case $U=0.8$ with $N=10^{4}$.
    Points are numerically obtained by averaging $20$ realizations.
    In panel (a), the symbols show the correlation function of momenta.
    The theoretical prediction, $\ln\tau/\tau$,
    is a better approximation than the simpler law $1/\tau$.
    Panel (b)  presents the diffusion of angles.
    Although the diffusion is normal,
    the straight line $\tau^{1.35}$ wrongly suggests that it is not.
    See text for explanations and details.
    Panel (c) shows the quantity
    $\sqrt{\sigma_{\theta}^{2}(\tau)/(\tau N^{2})}$ as a function of
    $\ln\tau$ to confirm the logarithmic correction of the diffusion.
    The straight line is a guide for the eyes.
    Finally, panel (d) presents the temporal evolution
    of the instantaneous exponent $\gamma$.
    The dashed line corresponds to  relation (\ref{eq:asymptotic-gamma}).}
  \label{fig:Cp-gaussian}
\end{figure}

Let us focus on the stable case $U=0.8$ with $N=10^{4}$.
The correlation function
obtained numerically, and shown in Fig.~\ref{fig:Cp-gaussian}(a), is
in good agreement with the theoretical prediction $(\ln \tau)/\tau$
in the long-time region $\tau>\tau_{1}=1$ if we accept the second
scaling of time as $\tau\to\tau/\tauscale$ with $\tauscale=0.2$. As
already mentioned in Sec.~\ref{sec:power-theory}, the second scaling
is not provided by the theory, while the asymptotic theoretical
estimate is out of applicability in the short time domain
$\tau<\tau_{1}$. We would like also to stress that the logarithmic
correction makes the prediction more precise rather than a simple
algebraic decay $1/\tau$.

The correlation function can thus be approximated as
\begin{equation}
  C_{p}(\tau) = \left\{
    \begin{array}{ll}
      C_{p}(0) & \quad \mbox{if}\ \tau<\tau_{1} \\
      &\\
      \dfrac{c\tauscale}{\tau} \ln \dfrac{\tau}{\tauscale}
      & \quad\mbox{if}\ \tau>\tau_{1}
    \end{array}
  \right. ,
\end{equation}
where the short time value has to be
$C_{p}(0)=\average{p^2(0)}_{N}=2K=0.6$, while $c=0.85$ is obtained
by a fitting procedure. This approximation of the correlation function and the
relation (\ref{eq:sigma_Cp}) leads to the following expression for
the diffusion
\begin{equation}
  \label{eq:gaussian-sigma-fitting}
  \dfrac{\sigma_{\theta}^{2}(\tau)}{N^{2}} = \left\{
    \begin{array}{ll}
      C_{p}(0)\tau^{2}, &
       \quad\mbox{if}\ \tau<\tau_{1} \\
      &\\
      2C_{p}(0)\left(\tau_{1}\tau-\dfrac{\tau_{1}^{2}}{2} \right)
      + c\tauscale
        \tau\left[ \left( \ln\dfrac{\tau}{\tauscale} \right)^{2}
          - \left( \ln\dfrac{\tau_{1}}{\tauscale} \right)^{2} \right]
      \\
        \hspace*{4.8em}- 2c\tauscale \left[
          \tau \left( \ln\dfrac{\tau}{\tauscale}-1 \right)
          - \tau_{1} \left( \ln\dfrac{\tau_{1}}{\tauscale}-1 \right) \right]
        &
         \quad\mbox{if}\ \tau >\tau_{1}
    \end{array}
  \right. .
\end{equation}
Figure~\ref{fig:Cp-gaussian}(b) presents the diffusion obtained
numerically. The two straight lines indicating the short- and
long-time regions shows a very good agreement. The diffusion seems
anomalous with an exponent $1.35$ in the long-time region.
Similarly, the instantaneous exponent $\gamma$ seems to converge
toward $1.35$ as shown by Fig.~\ref{fig:Cp-gaussian}(d). However,
these observations are not accurate, and only due to a long
transient, induced by the logarithmic correction. Diffusion is
essentially proportional to the time $\tau$, and hence must be
normal in the asymptotic time region. Expression
(\ref{eq:gaussian-sigma-fitting}) provides the asymptotic form of
the instantaneous exponent
\begin{equation}
  \label{eq:asymptotic-gamma}
  \gamma = 1 + \dfrac{2}{\ln(\tau/\tauscale)}.
\end{equation}
This prediction is in good agreement with numerical results as
attested by Fig.~\ref{fig:Cp-gaussian}(d). In the limit of
$\tau\to\infty$, the exponent $\gamma$ goes logarithmically toward
unity, and we therefore conclude that diffusion is normal although a
long transient time is necessary to observe it. This is an excellent
illustration of the difficulty to get reliable numerical estimates
for the diffusion exponent. Such a case explains {\em a posteriori}
the reason of previous disagreement~\cite{latora-99,yamaguchi-03}.

As predicted by Table~\ref{tab:Cp}, the logarithmic correction of
the correlation function yields a logarithmic correction of the
diffusion, so that its asymptotic form should be $
\sigma_{\theta}^{2}(\tau)/N^{2} \sim \tau(\ln\tau)^{2}$.
Figure~\ref{fig:Cp-gaussian}(c) confirms this prediction by plotting
$\sqrt{\sigma_{\theta}^{2}(\tau)/(\tau N^{2})}$ as a function of
$\ln\tau$: one gets a linear behavior in the long time region
$\tau>1$. We thus have confirmed the existence of weak anomalous
diffusion, i.e. normal diffusion with logarithmic corrections.

Let us return to the origin of discrepancies for
\textcolor{red}{$\alpha$ and} $\gamma$,
discussed at the end of the previous section
for the power-law tails. It seems natural to exclude the possibility (a),
lack of samples, since the same number of orbits, 20, has been used
in the case $N=10^{4}$, for both the power-tails and the gaussians,
while the latter case agrees extremely well with the theoretical
predictions, even including the logarithmic correction. This
excellent agreement comes from the absence of any breaking of
theoretical assumptions, since the state is at equilibrium and
stationary accordingly. Consequently, we can consider that the
possibility (b), lack of stationarity, explains the discrepancies of
$\alpha$ and $\gamma$ for power-law tails.

\section{Summary}
\label{sec:summary}

We have numerically confirmed the theoretical predictions proposed
in Ref.~\cite{Bouchet-Dauxois} for initial distributions with
power-law or gaussian tails: correlation function and diffusion are
in good agreement with numerical results.  Diffusion is indeed {\em
anomalous superdiffusion} in the case of power-law tails, while {\em
normal} when gaussian. In the latter case, the system is at
equilibrium, but the diffusion exponent shows a logarithmically slow
convergence to unity due to a logarithmic correction of the
correlation function. This long transient time to observe normal
diffusion, even for gaussian distribution and at equilibrium,
suggests that one should be very careful to decide whether diffusion
is anomalous or not~\cite{Correlation,Correlationb,Correlationc}.

For the power-law tails initial distribution, the numerically
obtained exponent of  diffusion is slightly different from the
theoretical prediction (few percents). As discussed above, this
discrepancy comes from the breaking of the stationary assumption.
The state is  only approximately stationary, explaining that the
theoretical predictions are not exact but only approximate. We
stress that in the limit of large $N$, these states become
stationary because their living times diverge much faster than $N$.
For the gaussian initial distribution, the state is in equilibrium
from the start, and stationary even with finite $N$: hence the
theoretical predictions agree extremely well with numerical results.

In addition, above numerical computations clarify two new points:
(i) the time region where the theory is applicable, (ii) the second
time scaling to fit the correlation function and the diffusion. Both
might depend on the degrees of freedom, but the latter, (ii),
appears to be not the case for the power-law tails. Obtaining the
dependence for the gaussian is a future work.

Finally, let us remark that the scenario of the relaxation described
in Refs.~\cite{yamaguchi-04, barre-06} is confirmed even for initial
distributions with power-law tails: this had never been tested
previously. The scenario claims that the system with long-range
interactions experiences first a violent relaxation, before the
so-called collisional relaxation which drives the system toward
Boltzmann's equilibrium. In the simulations reported here,
non-stationary and stable stationary states correspond respectively
to the violent and the collisional relaxations. One might also
remark that distributions with power-law tails induce
quasi-stationary states above the dynamical critical energy, while
being not a member of $q$-distributions~\cite{tsallis-88}. The
latter might be a sufficient condition of QSS, but is definitely not
a necessary condition.  To conclude let us remark that if the
results discussed here concerns the simple HMF model, let us mention
that it is believed to be general for long-range interacting
systems~\cite{Chavanis_Lemou,relatxationtest}.

\vspace*{1em}
\noindent
{\bf\large Acknowledgment}\\
YYY has been supported
by the Ministry of Education, Science, Sports and Culture,
Grant-in-Aid for Young Scientists (B), 16740223, 2006.

\end{document}